\documentclass[fleqn,usenatbib]{mnras}

\usepackage{newtxtext,newtxmath}
\usepackage[T1]{fontenc}

\usepackage{graphicx}	% Including figure files
\usepackage{amsmath}	% Advanced maths commands
\title[Periodicity of highly irregularly sampled light-curves]{Detecting the periodicity of highly irregularly sampled light-curves with Gaussian processes: the case of SDSS\,J025214.67-002813.7}

\author[Covino et al.] {
Stefano Covino,$^{1}$\thanks{E-mail: stefano.covino@inaf.it}
Felipe Tobar,$^2$
Aldo Treves,$^{1,3}$
\\
$^{1}$INAF / Brera Astronomical Observatory, via Bianchi 46, 23807, Merate (LC), Italy\\
$^{2}$Initiative for Data \& Artificial Intelligence, Universidad de Chile, Santiago de Chile, Chile.\\
$^{3}$Universit\`a degli Studi dell'Insubria, Via Valleggio 11, 22100 Como, Italy
}

\date{Accepted XXX. Received YYY; in original form ZZZ}

\pubyear{2021}

\begin{document}
\label{firstpage}
\pagerange{\pageref{firstpage}--\pageref{lastpage}}
\maketitle

\begin{abstract}
Based on a 20-year-long multiband observation of its light-curve, it was conjectured that the quasar SDSS\,J025214.67-002813.7 has a periodicity of $\sim 4.4$\,years. These observations were acquired at a highly irregular sampling rate and feature long intervals of missing data. In this setting, the inference over the light-curve's spectral content requires, in addition to classic Fourier methods, a proper model of the probability distribution of the missing observations. In this article, we address the detection of the periodicity of a light-curve from partial and irregularly-sampled observations using Gaussian processes, a Bayesian nonparametric model for time series. This methodology allows us to evaluate the veracity of the claimed periodicity of the abovementioned quasar and also to estimate its power spectral density. Our main contribution is the confirmation that considering periodic component definitely improves the modeling of the data, although being the source originally selected by a large sample of objects, the possibility that this is a chance result cannot be ruled out.
\end{abstract}


%%%%%%%%%%%%%%%%%%%%%%%%%%%%%%%%%%%%%%%%%%%%%%%%%%

%%%%%%%%%%%%%%%%% BODY OF PAPER %%%%%%%%%%%%%%%%%%

\section{Introduction}

Since the seminal paper of \citet{Begelmanetal1980}, the search of Binary Supermassive Black Holes (BSBH) has attracted  the interest of a vast group of astronomers; this is because BSBH are regarded as the parent population of the strongest gravitational wave (GW) signals, due to the final merging of the two black holes. Though such events have not been detected yet, current developments associated to, e.g., i) the discovery of GW bursts associated to collapsed star merging detected by the LIGO/Virgo Collaboration \citep{Abbottetal2020}, ii) the advancement of detection techniques through the Pulsar Timing Array (PTA) experiment \citep{Verbiestetal2021}, and iii) future space borne instrumentation like LISA \citep{Vitale2014,Sesana2021}, have strongly increased the development of the field in recent years.

Among the various approaches to finding  BSBH, a promising possibility has been that of detecting periodicities in the light curves of Active Galactic Nuclei (AGN). In particular, it is known that if the mass of the holes are of the order of $10^8$\,$M_\odot$ and the period is somehow related to the orbital one, with $\sim 0.01$\,pc separations, the periodicity should be of the order of years. Two main types of AGN have been considered thus far: Blazars (i.e. AGN dominated by a relativistic jet) and Quasars. Blazars largely populate the  gamma-ray sky, and most efforts related to the detection of their periodicities have been performed examining data from the Fermi Gamma-Ray Space Telescope \citep{Atwoodetal2009}, which covers the entire sky every three hours and has been collecting data for $\sim 11$\,years. For Quasars, main observational data are in the form of large collections of optical frames partly obtained from blind searches of transients. Data sources that are worth noticing in this regard are the Catalina Real-Time Transient Survey \citep{Djorgovskietal2011} and the Palomar Transient Factory \citep{Lawetal2009}.

Though there are thousands of Fermi gamma-ray blazars, those which are bright enough for constructing a light curve suitable for a temporal analysis are only a dozen. Careful examination of these blazars, performed by several researchers, has been possible due to the free availability of the data \citep[e.g. ][and references therein]{Covinoetal2019}. In general, the search of a period of about 1\,year is a challenging endeavour because: 1) due to uneven sampling, the required observation time needs to be about 10 times larger than the target period; 2) blazars are strongly variable sources, and the signal should be searched above a frequency dependent (red) noise; 3) the appearance of periodicity could be \emph{episodical}, that is, the signal may be identifiable on a time scale of a few years, but not of a decade or longer. However, one advantage of the Fermi light curves is that the sampling is homogeneous, without seasonal gaps.

Various methods have been used for the periodicity search. Some are more apt for sinusoidal signals, whereas others are essentially independent of the shape of signal, and often, unsurprisingly, results are widely distinct under the consideration of different methods \citep{Rieger2019}. The resulting picture is therefore still controversial. Arguably, the only source where the hypothesis of periodicity is supported by sufficient evidence is PG\,1553+113. The period $T \sim 2.2$\,years in the Fermi data was first proposed by \citet{Ackermannetal2015}, and then confirmed by \citet{Tavanietal2018}. Note, however, that its significance was challenged by \citet{AitBenkhalietal2019} and \citet{Covinoetal2019}, essentially due to the short number of epochs covered by the Fermi observations.

As for quasars, more than a hundred periodic candidates have been proposed in the last decade \citep[e.g.][]{Grahametal2015,Charisietal2016}, but the significance of some of these periods has been challenged because of the incorrect treatment of the multiple-trial effect due to the large number of considered sources \citep{Vaughanetal2016}. Finally, we have also to quote the case of OJ\,287, which is a Blazar, but its periodicity $T \sim 12$\,yr was identified from optical data using a century of archived plates \citep{Carrascoetal1985,Deyetal2019}. In this case, some concerns on the significance of the periodicity have also been raised \citep[e.g.][]{Stothers&Sillanpaa1997,Butuzova&Pushkarev2021}.

The methods for detecting the periodicity in blazars and quasars have proceeded practically independently, with little interactions  between the communities cultivating the two fields. See, however, \citet{Holgadoetal2018}, where the contributions of the two populations to the GW background, constrained by PTA observations, are compared and \citet{Charisietal2021} for future perspective of the field.

Recently, \citet{Chenetal2020} presented a systematic search of periodicity of 625 quasars with a median redshift $\sim 1.8$ and with multi-band photometry available for a time scale of $\sim 20$\,years. They found 5 candidates with periodicity of 3-5\,years. \citet{Liaoetal2021}, using the same dataset, considered very carefully the most interesting case, SDSS\,J025214.67-002813.7 (z=1.53), discussing the significance of the proposed $\sim 4.4$\,year ($\sim$ 1607\,days). \citet{Chenetal2021} focused on the possible physical interpretation for the identified periodicity.

In this paper we re-examine the available optical data for SDSS\,J025214.67-002813.7 (hereinfater J0252) using the same dataset as in \citet{Liaoetal2021}. Our analysis is carried out modeling the J0252 data by means of Gaussian processes (GPs) extending the procedure proposed in \citet{Covinoetal2020} for the case of blazars. The main goal of our work is to assess the statistical significance of the proposed periodicity and to determine the Power Spectral Density (PDS) of the target by means of a Bayesian non-parametric analysis.

Data about J0252 are described in Section\,\ref{sec:data}, the GP-based analysis methods are shortly described in Section\,\ref{sec:meth}. Results are reported in Section\,\ref{sec:res} and a discussion is developed in Section\,\ref{sec:disc}. Finally, conclusions are given in Section\,\ref{sec:concl}. Software libraries used for the computations carried out in this paper are listed in Appendix\,\ref{ap:soft}.

\section{SDSS\,J025214.67-002813.7}
\label{sec:data}

\begin{figure*}
\includegraphics[width=\textwidth]{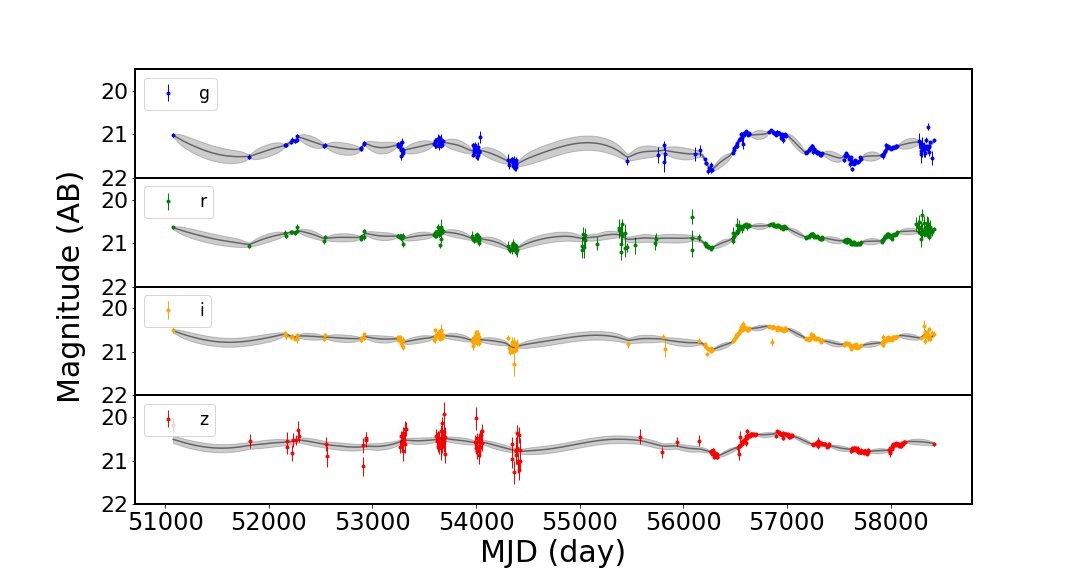}
\caption{$griz$ photometry for J0252 from \citet{Liaoetal2021}. Only points with 1$\sigma$ magnitude error lower than 0.3\,mag are plotted for clarity. The photometry covers about 20\,years although with seasonal gaps and interruptions and with an irregular sampling. Superposed to the four bands we also plot the best fit multi-band GP model and the 1$\sigma$ uncertainty, as discussed in Sect.\,\ref{sec:res}}.
\label{fig:lc}
\end{figure*}

J0252 was selected from a set of 625 quasars with spectroscopic identification \citep{Liaoetal2021}. For these objects, multi-band $griz$ photometry was available from the Dark Energy Survey \citep[DES, ][]{DES2016} and from the Sloan Digital Sky Survey \citep[SDSS, ][]{Ivezicetal2007} covering a range of about 20\,years of observations, with irregular samplings, gaps, etc. We refer the reader to \citet{Liaoetal2021} for any detail about data analysis. The light curves in the $griz$ are shown in Fig.\,\ref{fig:lc}. The median spacing is about 5-6 days, although the longest interruptions lasted more than about 1000\,days for all the light curves. In total the dataset consists of about 230-280 epochs.

\citet{Liaoetal2021} carried out a detailed analysis of the optical light-curves of J0252 and reported an interesting periodicity at $P = 1607 \pm 7$\,days with a significance of 99.95\% in the $g$ band, moreover the significance is larger than 99.43\% in any band. Given the length of the monitoring this implies that less than 5 complete cycles of variations have been covered. The short duration light-curve, in terms of the time scale under analysis, makes any periodicity claim intrinsically delicate. \citet{Liaoetal2021} identified the periodicity applying different analysis tools. A spectral analysis was carried out by the Lomb-Scargle algorithm \citep{Lomb1976,Scargle1982,Zechmeister&Kurster2009,VanderPlas2018} and the significance of the maximum power in the derived periodogram compared to the null hypothesis that it is due to noise was assessed by simulating a large number of artificial light curves with the same sampling of the original one. The analyses were also carried out in the temporal domain, fitting a sine curve or physically motivated models to the data, and modeling the auto-correlation function (ACF) of the data as a damped periodic oscillation \citep{Alexander1997,Grahametal2015}. J0252 turned out to be the most interesting case (i.e. with the lowest false alert probability) in a small set of five objects showing periodicities larger than 99.74\% in at least one band. The significance of the periodicity appears substantially stronger if all the bands are analysed together.

\begin{figure*}
\includegraphics[width=\textwidth]{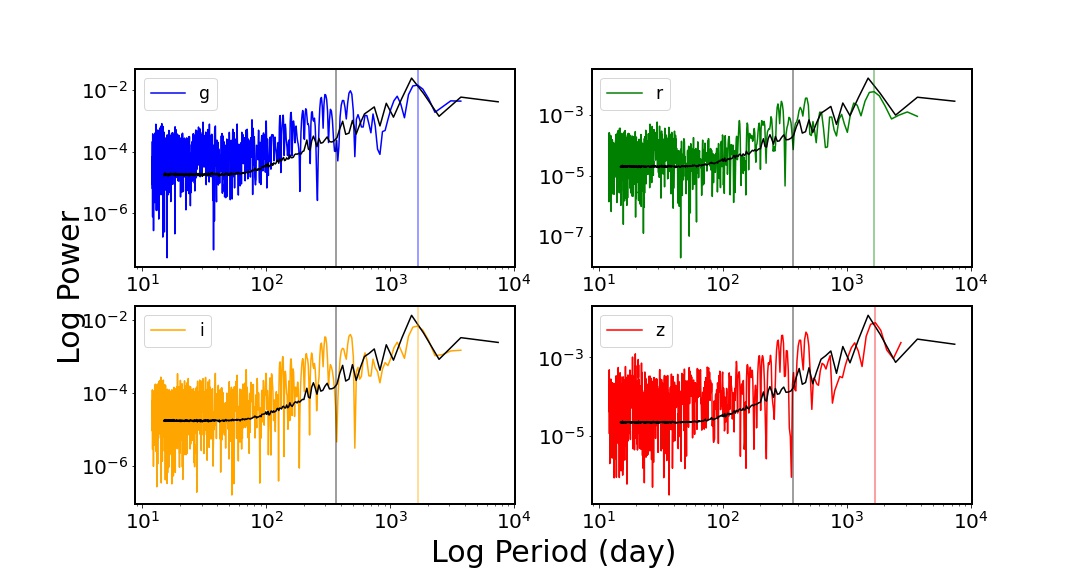}
\caption{Lomb-Scargle periodograms for the $griz$ bands of J0252. The colors lines are the actual periodograms while the black line is the PSD computed sampling at the maximum of the posterior distribution of the} the best-fit GP model (see Sect.\,\ref{sec:res}). The vertical lines indicate the maximum of the periodograms and the one year period.
\label{fig:LSplots}
\end{figure*}

In Fig.\,\ref{fig:LSplots} we show the Lomb-Scargle periodograms for the $griz$ bands, i.e. the convolution of the periodogram of the observed window together with that of the data \citep[e.g.,][]{vanderKlis1989}. The maxima of the periodograms are of course consistent with those reported in \citet{Liaoetal2021} and very similar in each band. However, the periodograms also show the typical signatures of correlated noise \citep[e.g.,][]{Vaughan2010}, with a flat region at shorter periods, where the periodograms are probably dominated by Poissonian noise, and a raising trend going to the longer (shorter) periods (frequencies). The presence of correlated noise and the short duration of the monitoring, compared to the proposed periodicities, make any claim about their significance somewhat ambiguous \citep[see also, e.g., ][]{Krishnanetal2021}. 

The Lomb-Scargle algorithm can also be obtained by a Bayesian analysis under the prior assumption that data periodicity are modeled by a sinusoidal functional form \citep{Jaynes&Bretthorst2003,Bretthorst2003,Mortieretal2015}. This scenario allows one to obtain the probability density function (PDF) for the frequency given the stated assumptions and the uncertainties associated to the maxima in the periodograms. In our case the PDF maxima and their $1\sigma$ uncertainties are: $1648 \pm 69$, $1657 \pm 97$, $1663 \pm 87$ and $1650 \pm 145$\,days, for the $griz$ bands, respectively.

\section{Gaussian processes}
\label{sec:meth}

Gaussian processes \citep[see][for a comprenhensive introduction]{Rasmussen&Williams2006} are probabilistic generative  models for time series \citep{Robertsetal2012}, which can be considered as the infinite-dimensional extension of the multivariate Normal distribution. As such, they provide a Bayesian nonparametric approach to smoothing, interpolation and forecasting \citep{Hogg&Villar2021}. Being defined on the real line rather than a finite, user-defined, discrete grid, GPs are particularly suited for (Bayesian) regression of irregularly-sampled data, possibly with large gaps, while providing sound error bars. 

The hyperparameters of the GP model are the mean and covariance functions. The former is usually set to zero or to linear/quadratic functions, while the latter is a  positive semidefinite function. The covariance \emph{kernel} is chosen based on a priori knowledge of the dynamic behavior of the data, for instance, the existence of periodicities or the requirement for differentiability. In addition to the well-known collection of off-the-shelf covariance functions \citep{Rasmussen&Williams2006},  the choice of a covariance function can rest upon the Wiener–Khinchin theorem \citep{Brockwell&Davis2016}, which states that ---in the stationary case--- the covariance kernel and the power spectral density (PSD) of the data  are Fourier pairs.

\subsection{Periodicity assessment}

GP techniques have been applied to identify possible periodicities in time series \citep{Robertsetal2012,Tobaretal2015,Durrandeetal2016,Littlefairetal2017,Angusetal2018,Tobar2018,Covinoetal2020,Coranietal2020,Zhangetal2021}. Here, we implement a procedure originally suggested by \citet{Littlefairetal2017} and then further developed by \citet{Covinoetal2020}. The idea is essentially to model the light curve of interest either with a stationary covariance function or with the same kernel but multiplied by a periodic (e.g., cosine) covariance function. Then, the problem of determining the presence of a periodic component hidden in the (often correlated) noise affecting an astronomical time series can be addressed by means of  Bayesian model selection \citep{Kass&Raftery1995, Jenkins2014, Andreon&Weaver2015, Trotta2017}.

\citet{Wilkins2019} and \citet{Griffithsetal2021} carried out extensive simulations verifying the capability of some of the most common kernel functions \citep[e.g. the squared exponential or radial basis, Mat\'ern and rational quadratic kernels, ][]{Rasmussen&Williams2006} to accurately  modeling the data PSD. Results show that, depending on the specific sampling pattern and gap duration, either the Mat\'ern and rational quadratic covariance functions typically outperform the squared exponential kernel. The Mat\'ern kernel family is characterized by a parameter, $\nu$, that drives the degree of ``smoothness" of the kernel. Time-series models corresponding to autoregressive processes of the first order are discrete time equivalents of GP models with Mat\'ern covariance functions with $\nu = 1/2$, which correspond, in one dimension, to the Ornstein-Uhlenbeck process \citep[e.g.][]{Robertsetal2012}. The Ornstein–Uhlenbeck process is frequently invoked in modeling the statistical behavior of AGN and blazars \citep[e.g.][]{Takataetal2018,Burdetal2021}. 

Given that the product and the sum of covariance kernels are again  legitimate (i.e., symmetric and positive definite) covariance kernels, we can build a periodic covariance function by multiplying a stationary kernel with the so-called ``cosine'' kernel \citep{Covinoetal2020}. Since a GP models are naturally suited for a Bayesian treatment, we can compute the probability favoring the more complex (i.e. periodic) model compared to a simpler one (non-periodic) by the Bayes factor. A remarkable by-product of this kind of analysis is that the PSD for the period can be easily obtained marginalizing out (i.e integrating) the posterior probability distribution of the parameters obtained from the regression.

%\subsection{Spectral estimation}
%\label{sec:specest}

%A different, although clearly strictly related, goal is based on the possibility to obtain a non-parametric modeling of the data PSD. Here, rather than trying to infer the importance of a periodic behavior we are more interested in deriving the actual shape of the PSD with all the information provided by the posterior distribution of parameters. Once again, GP regression is applied more for obtaining information about the covariance function and then about the PSD, rather then for fitting the light curve. This can become particularly challenging when only partial and noisy observations of the signal are available, where current methods typically fail to handle the uncertainties appropriately.

Spectral estimation by means of GP regression is an active research subject, which had a considerable boost after the seminal paper by \citet{Wilson&PrescottAdams2013} where the PSD is modeled by a Gaussian mixture in the Fourier domain. This way, the resulting covariance function (via the Wiener–Khinchin theorem) is a \emph{spectral mixture} of sub-kernels that allows one to approximate any stationary covariance kernel to arbitrary precision, given enough mixture components. The multioutput extension of the spectral mixture kernel was developed by \citet{Parra&Tobar2017,deWolffetal2020}. Other approaches have also been proposed. In the recent astrophysical literature, a mixture of the {\it celerite} covariance functions \citep{Foreman-Mackeyetal2017}, a family of physically motivated kernels developed also for allowing a very fast computation, was applied \citep{Yangetal2021} to derive the PSD from {\it Fermi}-LAT observations for a sample of blazars with published claims of quasi-periodicities. Another method, termed Bayesian nonparametric spectral estimation (BNSE), was proposed by \citet{Tobar2018}; BNSE is the limit of the well known Lomb-Scargle algorithm when an infinite number of components is considered with a Gaussian prior over the weights.

%We refer the reader interested in the full mathematical details of the algorithm to \citet{Parra&Tobar2017}, \citet{Tobar2018} and \citet{deWolffetal2020}. The idea is essentially to build a joint probabilistic model for the latent signal, a windowed version of the signal for which the Fourier transform exists, the closed-form posterior distribution of the windowed-signal spectrum, and finally the closed-form posterior power spectral density. Spectral estimation then becomes an exact inference problem.  

\section{Results}
\label{sec:res}

\subsection{Periodic vs non-periodic kernels}

We first modeled the light curves with the standard stationary kernels mentioned in Sect.\ref{sec:meth}, i.e., the Square Exponential (SE), the Mat\'ern with  $\nu = \frac{1}{2}$ aka absolute exponential (AE) and the rational quadratic (RQ) \citep{Rasmussen&Williams2006}. For all kernels, we denote $r = (t_i-t_j)$ the temporal difference between data points.

The SE kernel is given by:
\begin{equation}
k_{r} = A \exp \left(-\frac{r^2}{2L^2} \right),
\label{eq:SE}
\end{equation}
where $A>0$ is the amplitude and $L>0$ is the lengthscale of the exponential decay. %The SE kernel is a special case of the more general Mat\'ern covariance function family that can be obtained with $\nu \to +\infty$. 
The AE kernel (or, in one dimension, the kernel of the Ornstein-Uhlenbeck process) is given by:
\begin{equation}
k_{r} = A \exp \left(-\frac{r}{L} \right).
\label{eq:AE}
\end{equation}
Lastly, the RQ kernel\footnote{Note that the analogous formula reported in \citet{Covinoetal2020} is shown with a typo. We thank the anonymous referee for having pointed it out.} is:
\begin{equation}
k_{r} = A \left[1 + \left(\frac{r^2}{2\alpha L^2} \right)\right]^{-\alpha},
\label{eq:RQ}
\end{equation}
with $\alpha>0$. This kernel can be obtained by a scale mixture (i.e. an infinite sum) of SE covariance functions with different characteristic lengthscales drawn from a {\tt gamma} distribution. The limit of the RQ covariance for $\alpha \to +\infty$ is, again, the SE covariance function.

\begin{table}
\centering
\begin{tabular}{c|c}
\hline
\hline
Hyper-parameter & prior \\
\hline 
$\ln A$ & {\it Uniform} [-20, 20] \\
$\ln L$ & {\it Uniform} [-20, 20] \\
$\ln \alpha$ & {\it Uniform} [-10, 10] \\
$\ln P$ & {\it Uniform} [$\ln(100), \ln(3000)$] \\
\hline
\end{tabular}
\caption{Prior information adopted for for analyses described in Sect.\,\ref{sec:res}. Priors are properly normalized for the computation of the Bayes factors.}
\label{tab:priors}
\end{table}

In our regression experiments, the mean function of the GPs was set to the (constant) empirical mean of the observations. We also adopted flat priors for the kernel hyper-parameters as reported in Table\,\ref{tab:priors}. We considered both maximum likelihood implemented via L-BFGS-B  \citep[][]{Byrdetal1995} and also integrated out the hyperparameters from the posterior density by a Markov Chain Monte Carlo \citep[MCMC,][]{Hogg&Foreman2018} based on the ``parallel-tempering ensemble" algorithm \citep{Foreman-Mackeyetal2013,Goggans&Chi2004,Vousdenetal2016}. We started the Markov chains from small Gaussian balls centered on the best fit values. Then, we \emph{thinned} the chain by discarding a fraction of the samples corresponding to a few times (typically 4 or 5) the auto-correlation length of the computed chains and we checked that a stationary distribution was reached \citep{Sharma2017}. In most cases the posterior distribution of the parameters (hereinafter just ``posterior") is single-peaked (unimodal) and model comparison could be quickly implemented by means of, e.g., the Bayesian Information Criterion \citep[BIC,][]{Schwarz1978}. However, in the general case of periodic behavior assessment, we are interested in deriving the posterior PDF of the periods and therefore we carried out model comparison by a full computation of the Bayes factors \citep{Kass&Raftery1995, Trotta2017}.

Unsurprisingly,  GP regression is very effective in modeling (i.e. interpolating) our datasets \citep[e.g.][]{Covinoetal2020}, yet Bayes factors show very different results for the adopted kernels (Table\,\ref{tab:stationarybf}) reflecting their different ability in describing the data covariance \citep[][]{Wilkins2019,Griffithsetal2021}. 

\begin{table}
\centering
\begin{tabular}{c|rrr}
\hline
\hline
band & BF$_{SE-RQ}$   & BF$_{SE-AE}$ & BF$_{AE-RQ}$  \\
     &   dB  & dB & dB \\
\hline 
$g$ & $53$ & $44$ & $09$ \\
$r$ & $55$ & $68$ & $-13$ \\
$i$ & $66$ & $117$ & $-51$\\
$z$ & $5$ & $-32$ & $36$ \\
\hline
\end{tabular}
\caption{Bayes factors (expressed as dB) computed for the light curves in all the four available bands for a stationary kernel (e.g. RQ or AE) over another (SE or AE).}
\label{tab:stationarybf}
\end{table}

Bayes factors computed from noisy data are themselves noisy \citep[][]{Jenkins2014,Joachimietal2021}, therefore, only very large values unambiguously identify the most suitable hypothesis, given the data and the prior assumptions. Quite interestingly, results are different for each band, likely due to the different sampling and long-term coverage. The $g$ and $z$ band data are better described by the RQ kernel, while the $r$ and $i$ band data by the "AE" kernel. The SE kernel is generally disfavored although for the $z$ band curve it behaves almost comparably to the RQ kernel. A possible interpretation of this behavior depends on the presence of meaningful correlation for very long lags that, in case, can more effectively modeled by the RQ kernel \citep[][]{Rasmussen&Williams2006}. The different results in different bands are probably due to the more or less effective coverage for long lags and the quality (i.e. the associated uncertainties) of the data.

Perhaps the simplest periodic covariance function is the cosine kernel:
\begin{equation}
k_{r} = A \cos (2 \pi r / P),
\label{eq:cos}
\end{equation}
where $A>0$ is the amplitude, $P$ is the period and $r$ is the temporal difference as defined above. Given that the product of two kernels is still a valid positive semidefinite kernel, we modeled our data with a kernel constructed by multiplying the the best performing kernel in the previous experiment (Table\,\ref{tab:stationarybf}) with the cosine kernel.
We also adopted a flat prior distribution for the logarithm of the period as reported in Table\,\ref{tab:priors}.

\begin{table}
\centering
\begin{tabular}{c|rr}
\hline
\hline
band & BF$_{RQ-CRQ}$   & BF$_{AE-CAE}$  \\
   & dB & db \\
\hline 
$g$ & $21$ & $-$  \\
$r$ & $-$ & $23$  \\
$i$ & $-$ & $16$ \\
$z$ & $11$ & $-$  \\
\hline
\end{tabular}
\caption{Bayes factors (expressed as dB) computed for the light curves in all the four available bands for a periodic kernel over the best-fit stationary one.}
\label{tab:periodicbf}
\end{table}

With this new kernel, we witnessed a relevant improvement as quantified by the Bayes factors shown in Table\,\ref{tab:periodicbf}. Bayes factors, for ease of visualization, can be directly converted into probabilities conditioned on the data \citep[e.g.,][]{Trotta2007,Covinoetal2020} obtaining factors of 99.2, 99.4, 97.6, and 92.1\% for the $g$, $r$, $i$ and $z$ bands, respectively. In addition, the square root of the variances associated to each band, the $A$ parameter in Eq.(s)\,\ref{eq:SE}-\ref{eq:cos}, can be interpreted as modulation amplitudes giving: $\sim0.23, \sim0.14, \sim0.14$ and $\sim0.13$\,mag, respectively. These values are close to those reported in \citet[][their Table\,2]{Liaoetal2021}, although derived following a different approach.
%$0.23^{+0.10}_{-0.04}, 0.14^{+0.03}_{-0.02}, 0.14^{+0.04}_{-0.02}$ and $0.13^{+0.02}_{-0.02}$\,mag, respectively}.

\begin{figure*}
\includegraphics[width=\textwidth]{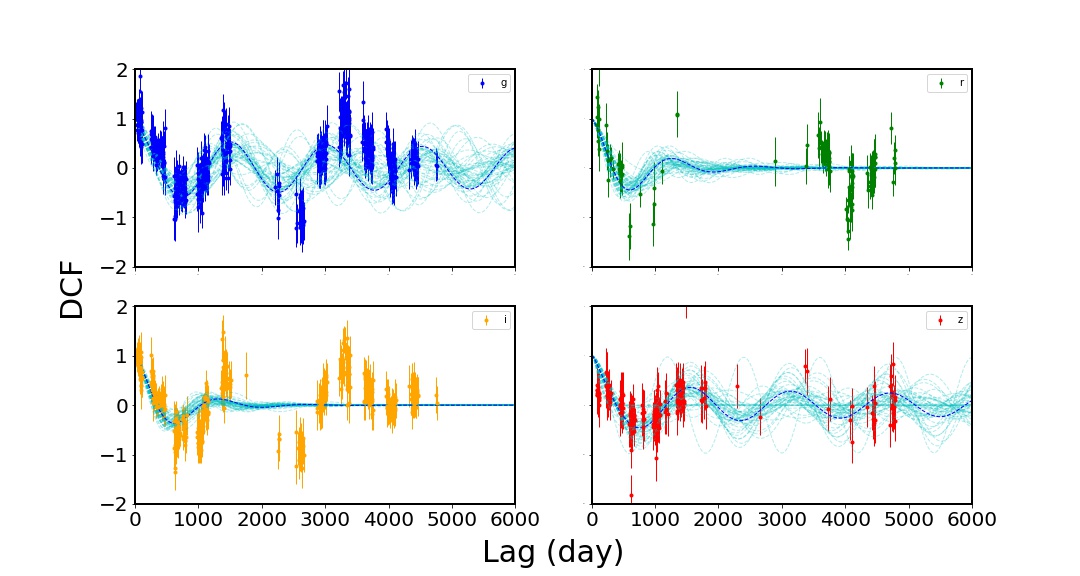}
\caption{DCF for the $griz$ bands of J0252. For clarity, only points with DCF errors lower than 0.4 are shown. The blue dashed line is the best-fit periodic covariance function for each band, and the cyan lines are 50 randomly chosen samples extracted from the posterior distribution of parameters.}
\label{fig:DCFplots}
\end{figure*}

A direct byproduct of performing GP regression is the conditional covariance of the process given the observations, which can be compared to the ACF of the data. Computing the ACF for data affected by a very irregular sampling and long gaps is not an easy task. We followed here the discrete correlation function (DCF) algorithm proposed by \citet{Edelson&Krolik1988}, although we stress that the DCF computed in this paper are shown only for illustration purposes.  Fig.\,\ref{fig:DCFplots} show the (rather) noisy DCF compared to the GP regression best-fit parameters. As mentioned already, the preference for the RQ or the AE kernel depends on the presence of correlation for long lags. The data, however, clearly show an oscillatory behavior, yet the highly-irregular sampling makes it difficult to assess whether we have a true periodicity or a simply a recurrent behavior with changing time scale.

\begin{figure}
\includegraphics[width=\columnwidth]{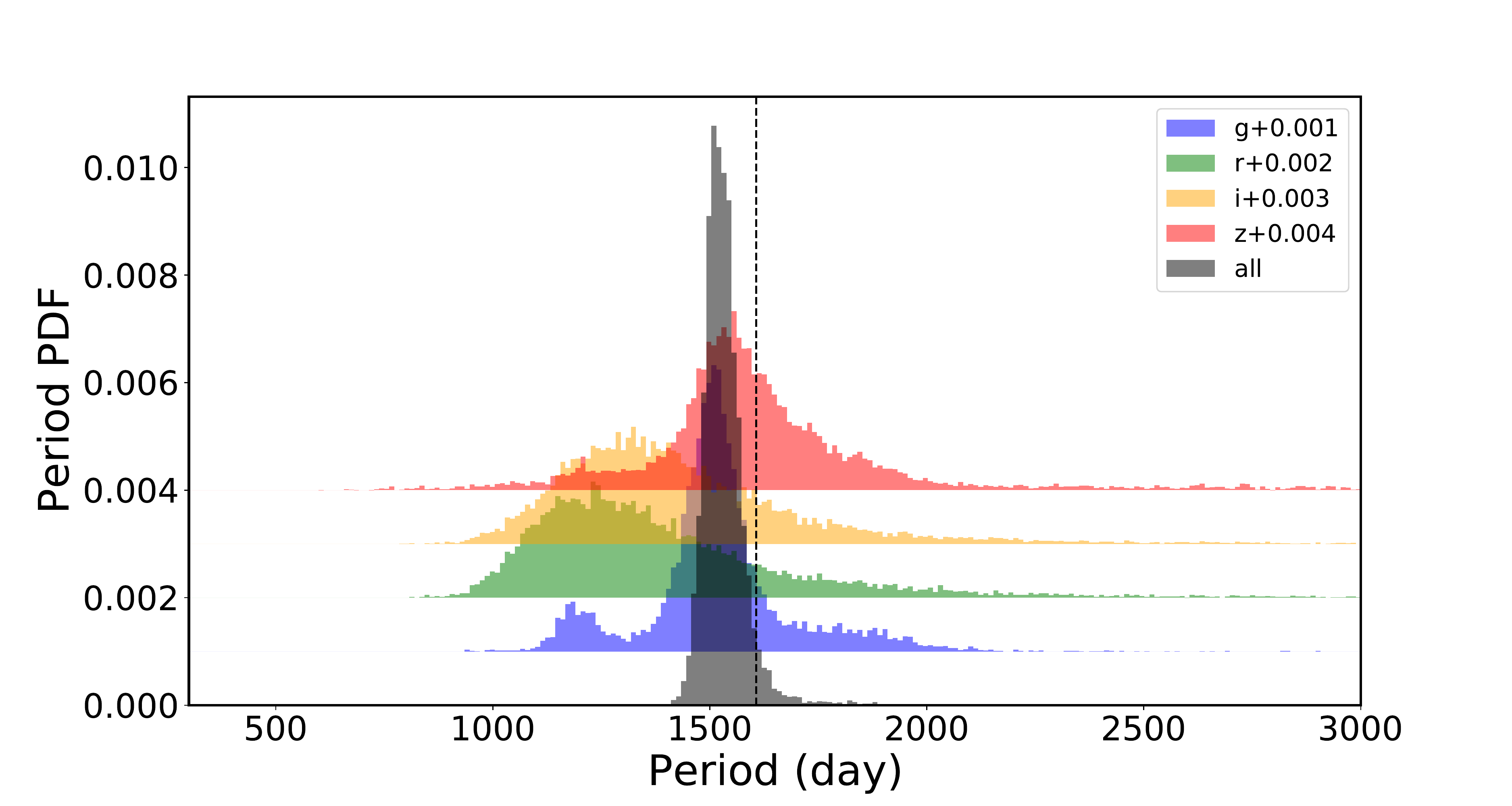}
\caption{PDF for the periods computed marginalizing the posterior density distributions for the $griz$ bands of J0252 and for the analysis with all the available bands together. The dashed vertical line identifies the period proposed in \citet{Liaoetal2021}. PDF for each band are artificially shifted for clarity by the amount reported in the legend. The most probable periods for the $griz$ bands are $1511^{+103}_{-109}$, $1317^{+323}_{-183}$, $1570^{+220}_{-133}$, and $1373^{+243}_{-215}$\,days, respectively. The most probable period for the analysis including all the available bands is $P = 1525^{+42}_{-33}$\,days.}
\label{fig:HistPer}
\end{figure}

Within the limits of our models, we can expect that the periods are better constrained when the RQ kernel is preferred, since the AE kernel cannot support correlation on long lags. This is indeed the case, as we can see in Fig.\,\ref{fig:HistPer}.
As expected, the PDF for the $g$ and partly for the $z$ bands are more peaked while the range of possible periods for the $r$ and $i$ bands are larger. Periods maximising the GP regression posterior PDF tend to be shorter that those maximising the LS periodograms. A similar behavior was also obtained by \citet{Liaoetal2021} (their Table\,2) comparing results from LS and time domain analyses.

Single-band study does not actually make the best use of all the available information. Having data collected in four different bands allows us to obtain a more solid inference modeling the $g, r, i$ and $z$ band data together and asking that the period be same at all frequencies. This is an important step that is getting more and more significant given the common availability of multi-band information from modern and future monitoring facilities \citep[see, e.g.,][]{Huijseetal2012,Vanderplas&Ivezic2015,Mondriketal2015,Saha&Vivas2017,Huijseetal2018,Hu&Tak2020,Elorrietaetal2021}. 

However, analyzing multi-band time-series requires some care since we may want to consider the, typically large, degree of correlation among the different bands. This is a basic topic in the GP literature \citep[][]{Pinheiro&Bates1996,Robertsetal2012,Osborneetal2012,Parra&Tobar2017} due to the clear interest in deriving information from multiple inputs in many sectors from research to industry \citep[][]{Robertsetal2012}.

In this work we have applied a simple technique known as ``{\it intrinsic model of coregionalization}" \citep[ICM, ][]{Bonillaetal2008,Alvarezetal2011} although many other, often more sophisticated, approaches are possible \citep[e.g.][]{vanderWilketal2020,deWolffetal2020}.
Assuming we have $P$ different input data sets (or bands in our case) the kernel for our problem can be defined as:
\begin{equation}
K_{\rm coreg} = k(x,x') \cdot B[i,j],    
\end{equation}
where $x$ and $x'$ are observations from any input data sets, $k(x,x')$ is a valid covariance function, as those introduced before, and $B$ is a $P \times P$ positive-definite matrix. $K_{\rm coreg}$, therefore, turns out to be:
\begin{equation}
K_{\rm coreg} =
\begin{bmatrix}
    B_{11} k(X_1,X_1) & \dots & B_{1P} k({X_1,X_P}) \\
    \vdots & \ddots & \vdots \\
    B_{P1} k(X_P,X_1) & \dots  & B_{PP} k(X_P,X_P) 
\end{bmatrix}.
\end{equation}

Each output (band, in our case) can be seen as linear combination of functions defined by the same covariance $k(x,x')$. The rank of the $B$ matrix defines the number of component in the linear combination. Therefore, an ICM is generated by the products of two covariance functions:
one that models the dependence between the outputs and one that models the input dependence \citep[][]{Alvarezetal2011}.
The restriction $B_{ij} = 0$ for $i \ne j$ describes the case with independent data for bands $i$ and $j$. 

ICM models can be, however, computationally demanding. In the high rank case, the number of hyper-parameters to be learned is (in our case) larger than 20, making the computation of the likelihood computationally expensive. Therefore, for computational purposes, and also because the rather high degree of correlation and superposition among the four available bands makes high rank coregionalized matrices unnecessary, we applied the simplest model with the lowest rank, thus modeling our four bands simultaneously with either the AE or the RQ covariance function. 

As expected, once the kernel hyper-parameters are learned using the whole dataset, the RQ kernel outperforms the AE, essentially because the former is better suited to model correlation for long lags. The introduction of a periodic component, now with the same period for all bands, is highly favoured by the computed Bayes factors (Table\,\ref{tab:allperbf}), the probability in favour of the periodic model is about 4.2$\sigma$. The PDF for the period exhibits a strong peak (Fig.\,\ref{fig:HistPer}) and yields a period of $P = 1525^{+42}_{-33}$\,days.

\begin{table}
\centering
\begin{tabular}{rr}
\hline
\hline
 BF$_{AE-RQ}$   & BF$_{RQ-CRQ}$  \\
   dB & db \\
\hline 
 $64$ & $45$  \\
\hline
\end{tabular}
\caption{Bayes factors (expressed as dB) computed for the light curves in all the four available bands considered together for a periodic kernel over the best-fit stationary one.}
\label{tab:allperbf}
\end{table}

As mentioned before, adopting a higher-rank formulation quickly makes the (Monte Carlo) computation of the likelihood (and the Evidence) impractical due to the large number of parameters. Popular and, for simple inferences, effective alternatives are available as the already introduced BIC, which is a proxy of the actual Bayes factor, i.e. $\log {\rm Evidence} \approx (-BIC/2)$ \citep[see, e.g. ][for pros and cons of the BIC for model inference]{Kass&Raftery1995,Raftery1999}. Else, more advanced techniques for approximate Evidence computation that scale efficiently with the number of parameters have also been developed  \citep[e.g. variational inference,][]{Opper&Archambeau2009,Bleietal2016,Cherief-Abdellatif2019}.

Finally, as discussed in Sect.\,\ref{sec:meth}, another by-product of our analysis is an estimate of the PSD of the data. Applying again here a rank one ICM model with the best-fit periodic covariance function, we can derive a PDF directly from the GP regression (Fig.\,\ref{fig:LSplots}). Given that the PSD can be computed for any needed sampling rate, it can be used for complex inferences or even fit as, e.g., in \citet{Vaughan2010} with any physically-driven model. More advanced GP based algorithms could also be applied, if needed, to derive the PSD with more general assumptions about the covariance of the data \citep[e.g.][]{deWolffetal2020}. 

\section{Discussion}
\label{sec:disc}

The analyses carried out in Sect.\,\ref{sec:res} show that a covariance function including a periodic component is supported by the data with a probability higher than 99\% for the $g$ and $r$ bands.
These improvements are of interest, although somehow less significant compared to the results reported by \citet{Liaoetal2021}. Results obtained through different techniques should be carefully analysed since their direct comparison can be occasionally misleading due to their different assumptions. In addition, we also stress that here we report probabilities supporting a periodic kernel with ``any period" within the adopted prior, since we are considering the period as a nuisance parameter and thus integrating it out from our posterior. With essentially single-peaked unimodal distributions it makes little difference, but this does not need to be generally the case \citep[see also, e.g. ][for a similar discussion]{Gregory&Loredo1992}. 

The availability of multiple bands allows us to carry out the analysis making use of all the information provided by the dataset, and the probability supporting a periodic kernel given the data becomes higher, corresponding to an improvement with respect the purely stationary channel-independent solution of about $4.2\sigma$. The best-fit period, $P = 1525^{+42}_{-33}$\,days, is shorter than those reported in \citet{Liaoetal2021} and those obtained by our LS analysis (Sect.\,\ref{sec:data}), although still roughly consistent given the uncertainties. This difference might indicate that the period is not stable during the monitoring, since LS analysis is essentially a change of basis, and data are simply described by the frequency content. A GP regression models a covariance function and the level of correlation for short and long lags affects how the best period describing the data is obtained. Shorter periods than those identified by a LS analysis were reported in \citet{Liaoetal2021} too (see their Table\,2).

In any case, the source considered in \citet{Liaoetal2021} and in this study was singled out by a sample of 625 analysed objects \citep{Chenetal2020}. This implies the reported probabilities have to be corrected for the so-called ``look elsewhere effect" \citep{Bayer&Seljak2020}, i.e., the probability to find a positive result across multiple (independent) trials. In our case the $\sim4.2\sigma$ probability supporting the periodic kernel becomes, after the multiple trial correction \citep{vanderKlis1989}, $\sim 2.4\sigma$. This is still an interesting figure, yet far from being conclusive, implying that the periodicity proposed for J0252 can be just a chance result is still a plausible interpretation. We mention in passing that the multiple trial correction \citep[see a discussion in ][]{Vaughan2013} is often neglected since it is occasionally left unrecorded or badly described how large is the original sample of studied sources before devoting specific attention to the most interesting cases. Unfortunately, it is a factor that can easily dominate the final false alert probability for a proposed period.

\section{Conclusions}
\label{sec:concl}

In this paper we have collected data for an AGN with a long multi-band optical monitoring, SDSS\,J025214.67-002813.7. This source was singled out by \citet{Liaoetal2021} from a larger sample of objects and, after an extensive analysis, they proposed a possible periodicity of $\sim 4.4$\,years. The available data cover a long period, a couple of decades, but are affected by a very irregular sampling and long gaps. We re-analyzed these data modeling the light curves by Gaussian processes, i.e. Bayesian nonparametric models for regression of time series . We analysed the single band data and by means of multi-output GP models the whole dataset at once. We can confirm that a periodic component slightly longer than $\sim 4$\,years improves the description of the data in all bands, and is also supported by the full analysis of the whole dataset. Nevertheless, given that the considered source was originally identified in a large sample of several hundreds sources, after a correction due to the sample size the false alert probability of the periodicity does not appear to be low enough to rule out the possibility this is just a chance result. 

\section*{Acknowledgements}

We thank the anonymous referee for her/his valuable and accurate comments. We acknowledge partial funding from Agenzia Spaziale Italiana-Istituto Nazionale di Astrofisica grant I/004/11/5. FT acknowledges ANID grants  Fondecyt-1210606, ACE210010, FB210005 \& FB0008.

%%%%%%%%%%%%%%%%%%%%%%%%%%%%%%%%%%%%%%%%%%%%%%%%%%
\section*{Data Availability}

The data considered in this paper were published in \citet{Liaoetal2021}.



\bibliographystyle{mnras}
\bibliography{QuasarQPO} % if your bibtex file is called example.bib

%%%%%%%%%%%%%%%%% APPENDICES %%%%%%%%%%%%%%%%%%%%%

\appendix

\section{Software packages}
\label{ap:soft}

We have developed software tools and used third-party libraries all developed with the {\tt python} language \citep{vanRossum1995} (v. 3.7-8-9)\footnote{http://www.python.org} with the usual set of scientific libraries ({\tt numpy} \citep{numpy} (v. 1.15.4)\footnote{http://www.numpy.org} and {\tt scipy} \citep{scipy} (1.10)\footnote{https://www.scipy.org}. The generalised LS algorithm we applied is part of the {\tt astropy} (v. 3.1.2)\footnote{http://www.astropy.org} suite \citep{astropy2013,astropy2018}. Non-linear optimization algorithms and numerical integration tools are provided by the {\tt minimize} and {\tt integrate} subpackages of {\tt scipy} library. MCMC algorithms are provided by the {\tt ptemcee}\footnote{https://github.com/willvousden/ptemcee} (v. 1.0.0) library \citep{Foreman-Mackeyetal2013, Vousdenetal2016}. GP analysis is carried out by the {\tt george} package (v. 0.3.1)\footnote{https://george.readthedocs.io/en/latest/} \citep{Ambikasaranetal2014} and by the {\t GPFlow} package (V. 2.2.1)\footnote{https://gpflow.readthedocs.io/en/master/index.html} \citep[][]{Matthewsetal2017}. Multi-output GP models have been implemented also with the {\tt MOGPTK} package (v. 0.2.5)\footnote{https://github.com/GAMES-UChile/mogptk} \citep[][]{deWolffetal2020}. The DCF is computed by the tools in the {\tt astroML} library\footnote{https://www.astroml.org/} (v. 0.4.1) \citep{VanderPlasetal2012,Ivezicetal2014}.
Plots are produced within the {\tt matplotlib} \citep{Hunter2007} (v. 3.0.2)\footnote{https://www.matplotlib.org} framework. Multidimensional projection plots were obtained with the {\tt corner} \citep{Foreman-Mackey2016} (v. 2.0.2)\footnote{https://corner.readthedocs.io/en/latest/} library. Some of the developed code was executed by means of the {\tt Google Colab} toolkit \citep[][]{Bisong2019}.
%%%%%%%%%%%%%%%%%%%%%%%%%%%%%%%%%%%%%%%%%%%%%%%%%%

% Don't change these lines
\bsp	% typesetting comment
\label{lastpage}
\end{document}